\documentclass[a4paper,twoside]{article}

\usepackage{epsfig}
\usepackage{subcaption}
\usepackage{calc}
\usepackage{amssymb}
\usepackage{amstext}
\usepackage{amsmath}
\usepackage{amsthm}
\usepackage{multicol}
\usepackage{pslatex}
\usepackage{apalike}
\usepackage{tabularx}
\usepackage[bottom]{footmisc}
\usepackage{cmap}                                 
\usepackage[english]{babel}   
\usepackage{tabularx}                             
\usepackage{booktabs}                             
\usepackage{rotating}                             
\usepackage{multirow}                             
\usepackage{amsmath}
\usepackage{hyperref}                             
\usepackage{flafter}                              
\usepackage{pdflscape}                            
\usepackage{hyphenat}                             
\usepackage[all]{hypcap}                          
\usepackage{url}                                  
\usepackage{enumitem}                             
\usepackage{graphicx}
\usepackage{iflang}

\usepackage{verbatim}

\usepackage{amsfonts}
\usepackage{mathrsfs}

\usepackage{graphicx}
\usepackage{caption}
\usepackage{colortbl}
\usepackage{pdfpages}
\usepackage{verbatim}
\usepackage{subcaption}
\usepackage{mwe}

\usepackage{placeins}
\usepackage{pythonhighlight}
\usepackage{listings}
\usepackage{algorithm}
\usepackage{algpseudocode}
\usepackage{tikz}
\usepackage{comment}

\usetikzlibrary{babel}
\usetikzlibrary{decorations.pathreplacing}
\usetikzlibrary{automata,arrows,positioning,calc}

\usepackage{SCITEPRESS}     

\begin{document}

\title{Learning to Participate through Trading of Reward Shares}

\author{\authorname{
Michael Kölle\sup{1}, 
Tim Matheis\sup{1}, 
Philipp Altmann\sup{1},
Kyrill Schmid\sup{1}
}
\affiliation{\sup{1}Institute of Informatics, LMU Munich, Oettingenstraße 67, Munich, Germany}
\email{\{michael.koelle, philipp.altmann, kyrill.schmid\}@ifi.lmu.de}
}


\abstract{
Enabling autonomous agents to act cooperatively is an important step to integrate artificial intelligence in our daily lives. While some methods seek to stimulate cooperation by letting agents give rewards to others, in this paper we propose a method inspired by the stock market, where agents have the opportunity to participate in other agents' returns by acquiring reward shares. Intuitively, an agent may learn to act according to the common interest when being directly affected by the other agents' rewards. The empirical results of the tested general-sum Markov games show that this mechanism promotes cooperative policies among independently trained agents in social dilemma situations. Moreover, as demonstrated in a temporally and spatially extended domain, participation can lead to the development of roles and the division of subtasks between the agents. 
}

\onecolumn \maketitle \normalsize \setcounter{footnote}{0} \vfill

\section{Introduction}
The field of cooperative AI seeks to explore methods which establish cooperative behavior among independent and autonomous agents \cite{dafoe2020open}. The ability to act cooperatively is a mandatory step in order to integrate artificial intelligence in our daily lives especially in applications where different decision makers interact like autonomous driving. Various breakthroughs in the field of single agent domains \cite{mnih2015human,silver2017mastering} have also led to the successful application of reinforcement learning in the field of multi-agent systems \cite{leibo2017multi,phan2018leveraging,vinyals2019grandmaster}. However, while purely cooperative scenarios, where all agents receive the same reward and thus pursue the same goal, can be addressed with centralized training techniques, this is not the case if agents have individual rewards and goals. Moreover, if agents share resources it is likely that undesired behaviors are learned, especially when resources are getting scarce \cite{leibo2017multi}. Independent optimization may lead to sub-optimal outcomes such as for the Prisoner's dilemma or public good games. In more complex games, the agents’ risk-aversion as well as information asymmetry additionally deteriorate the likelihood of a desired outcome \cite{bargaining}. 

\begin{figure}
     \subfloat[PD without particpation \label{fig:prisoners-pd1}]{%
       \includegraphics[width=0.48\linewidth]{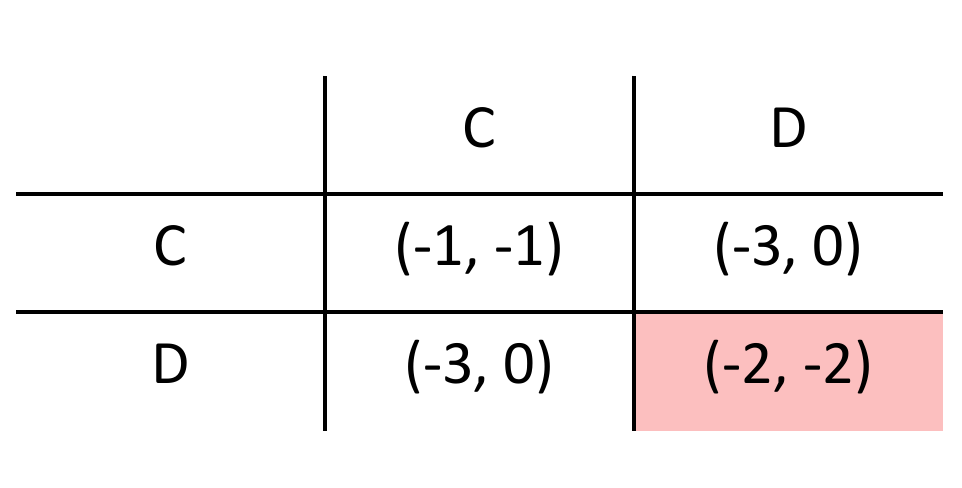}
     }
     \hfill
     \subfloat[PD with 50\% particpation \label{fig:prisoners-pd2}]{%
       \includegraphics[width=0.48\linewidth]{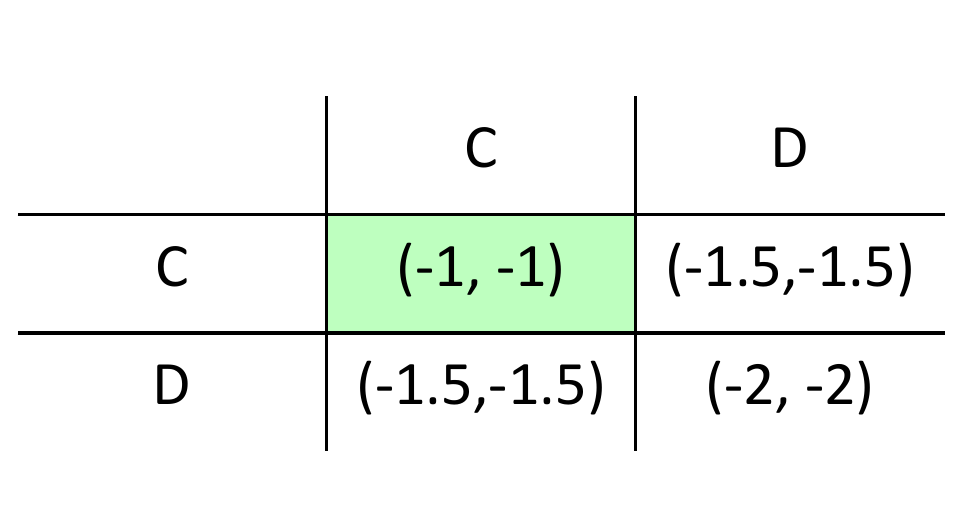}
     }
     \caption{By enabling agents to trade shares in their payoffs, a socially optimal outcome may be achieved.}
     \label{fig:prisoners}
\end{figure}

In recent years, various approaches have been proposed to promote cooperation among independent agents, such as learning proven game theoretic strategies like tit-for-tat \cite{lerer2017maintaining}, the possibility for agents to incentivize each other to be more cooperative \cite{schmid2018action,lupu2020gifting,lio}, or the integration of markets to let agents trade for increased overall welfare \cite{ijcai2021-54}. 
In this work, we adopt the market concept in order to generate increased cooperation between independent decision makers. More specifically, we propose a method that allows agents to trade shares of their own rewards. In the presence of such a participation market, we argue that a better equilibrium can be reached by letting agents directly participate in other agents' rewards. A socially optimal equilibrium may be established due to the direct incorporation of all global rewards instead of only incorporating individual rewards. Enabling agents to trade their shares at a fair price, while at the same time creating a trading path that in fact leads to a beneficial distribution of shares, represents the hardest challenge when implementing this model.

To motivate the effectiveness of participation, consider the Prisoner's dilemma (PD) as depicted in \autoref{fig:prisoners}. In the standard version of the Prisoner's dilemma, each agent receives its individual reward. For two rational decision makers, the only Nash equilibrium lies in mutual defection (DD), which is socially undesirable. However, if we introduce the ability of reward trading and each agent holds shares in the other agent's return (say $50$\% participation), then the overall dynamic changes and mutual cooperation becomes a dominant strategy. 
In order to train agents to participate, we apply different methods of reinforcement learning, so that the agents autonomously find strategies. Thus, participation in other agents' returns is realized over the course of many training episodes. To that end, different variants of the \emph{market for participation} are tested in this work, which differ in the way the participation mechanism is implemented. More specifically, the following contributions are made:
\begin{itemize}
    \item A theoretical motivation is given why market participation is beneficial.
    \item The participation mechanism is empirically evaluated in the Prisoner's dilemma as well as in a complex multi-agent scenario, called the clean-up game \cite{lio}.
\end{itemize}
All code for the experiments can be found here \footnote{\url{https://github.com/TimMatheis/Learning-to-participate}}.






\section{Related Work}\label{sec:relatedwork}
Despite the increasing success of reinforcement learning on an expanding set of tasks, most effort has been devoted to single-agent environments as well as fully cooperative multi-agent environments \cite{mnih2015human,silver2017mastering,vinyals2019grandmaster,berner2019dota}. However, with multiple agents involved their goals are often not (perfectly) aligned, which renders centralized training techniques in general unfeasible. The drawback of fully decentralized models is that agents focus only on their individual rewards, which therefore might result in undesirable collective performance especially in situations of social dilemmas or with common pool resources \cite{leibo2017multi,perolat2017multi}.

Previously, one way of tackling this problem has been to give independent agents intrinsic rewards \cite{eccles2019learning,hughes2018inequity,wang2018evolving}. The concept of intrinsic rewards draws from concepts in behavioral economics such as altruistic behavior, reciprocity, inequity aversion, social influence, and peer evaluations. These intrinsic rewards are usually either predefined or they evolve based on the other agents' performance over the game. Other works suggest that reward mechanisms \cite{lio} or penalty mechanisms \cite{kyrill} may lead to cooperation in sequential social dilemmas. The literature distinguishes between selective incentives and sanctioning mechanisms which incentivize cooperative behavior in social dilemmas \cite{kollock1998social}. Selective incentives describe methods that attempt to positively promote cooperation. For instance, this could occur by giving monetary rewards to reduce the consumption of common pool goods, such as water or electricity \cite{maki1978time}. Contrarily, penalties could be a method to reduce defective behavior. In fact, experiments with humans suggest that penalties are effective in reducing defective behavior \cite{komorita1987cooperative}. 

In \textit{LIO} \cite{lio}, a reward-giver’s incentive function is learned on the same timescale as policy learning. Adding an incentive function is a deviation from classical reinforcement learning, where the reward function is the exclusive property of the environment, and is only altered by external factors. As shown by empirical research \cite{lupu2020gifting}, augmenting an agent’s action space with a “give-reward” action can improve cooperation during certain training phases. Through opponent shaping, an agent can influence the learning update of other agents for its own benefit. A different attempt at opponent shaping is to account for the impact of one agent's policy on the anticipated parameter update of the other agents \cite{lola}. Through this additional learning component in \textit{LOLA}, strategies like tit-for-tat can emerge in the iterated Prisoner's dilemma, whereby cooperation can be maintained.


Other work in the field suggests markets as vehicles for cooperativeness \cite{schmid2018action,ijcai2021-54}. Usually, agents only receive their individual rewards. As they are not affected by the other agents' individual rewards, they only act in their own interest. However, by only receiving individual rewards, the agents are exposed to substantial risk. According to economic theory, it is usually beneficial to be diversified. In portfolio theory, there is a common agreement that diversification increases expected returns. Although people do not seem to diversify enough, which is called the \textit{diversification puzzle} \cite{statman2004diversification}, a rational agent should be perfectly diversified and should only hold a combination of the market portfolio and a risk-free asset (such as a safe government bond). When applying this to games such as the Prisoner's dilemma, the agents should be interested in receiving a combination of their individual rewards and the other agents' individual rewards to minimize their risk exposure. 
\section{Learning to Participate}
\label{sec:participate}

Multi-agent systems consist of multiple agents that share a common environment. An agent is an autonomous entity with two main capabilities: perceiving and acting. The perception of the current state of the environment allows the agent to choose an appropriate action out of an action set. The chosen action depends on an agent's policy. Reinforcement learning methods are often applied to teach an agent a good policy in multi-agent reinforcement learning.

Since cooperative strategies can be difficult to find and maintain, we suggest a participation mechanism. The idea is to let agents participate in other agents' environmental rewards directly in order to align their formerly conflicting goals. In the following, we use the trading of participation shares as an instrument to make cooperation possible. If the agents are willing to hold a significant amount of shares in all agents' rewards, they may act in the ``society's interest''. Namely, the difference of the individual interest of a single agent and the common interest of the collective of all agents could vanish. In our implementation, two agents only trade shares whenever both choose to increase or decrease the amount of their own shares. In the initial state, agents hold 100\% of their own shares. If the agents want to be perfectly diversified, each agent could have $\frac{100\%}{n}$ shares, with $n$ denoting the number of agents in the environment, of every agent's rewards after some trading steps.

\section{Experiments}
\label{sec:exp}
In this section, we examine our theoretical considerations in the iterated prisoner's dilemma and the clean-up game experimentally.

\subsection{The Iterated Prisoner's Dilemma}
\label{sec:ipd}
The prisoner’s dilemma can be thought of as the action of two burglars. When they get caught, they can decide between $a_0 = admitting$ and $a_1 = not admitting$ a crime. If both admit the crime, they receive a high punishment. If none of them admits the crime, they only receive a low punishment. 
The dilemma evolves from the case in which only one of them admits the crime. Then, the admitter is not punished due to its status as a principal witness, whereas the denier receives a very high punishment. 
Admitting is the defective ($D$) action and not admitting is the cooperative ($C$) action in \autoref{fig:prisoners}. By definition, none of the agents can put itself in a better position by changing its strategy in a Nash equilibrium. Agent 1 knows that agent 2 can defect and cooperate. When agent 2 defects, agent 1 is better off by defecting as well. If agent 2 cooperates instead, agent 1 is again better off by defecting. Thus, agent 1 should defect in any case, which makes defecting a dominant strategy in a one-shot game. By symmetry, agent 2 faces the same problem and should also defect. The tragedy is that this outcome is not desirable, as mutual cooperation leads to a better payoff for both agents. When the game is iterated multiple times, defecting is in theory not necessarily dominant. 
Although the agents could defect in every iteration based on backward induction, 
the agents could develop strategies to incentivize cooperation.

\subsection{Participation in the Iterated Prisoner's Dilemma}
\label{sec:particip}

For testing different implementations of the iterated prisoner's dilemma, we use an actor-critic method. We adapt Algorithm 1 from \textit{LIO} (\cite{lio} page 5) by replacing the incentive function and its parameter $\eta$ with a greater action space during the whole episode or a preliminary trade action. Hence, either the actions specify an environment action as well as a trade action during the whole episode, or there is one additional step at the start of each episode in which the participation is determined, or both. 

We use the same fully-connected neural network for function approximation as \textit{LIO}. However, we only use the policy network, as we do not make use of the incentive function, for which \textit{LIO} uses another neural network. The policy network has a softmax output for discrete actions in all environments. For all experiments, we use the same  architecture and the same hyperparameters: $\beta = 0.1$, $\epsilon_{\text{start}} = 1.0$, $\epsilon_{\text{end}} = 0.01$, $\alpha_{\theta} = 1.00E-03$. As the agents do not participate in the other agent's rewards at the start, \textit{max steps} is set to 40 instead of 5 in the implementations with trading. Hence, they have enough steps for trading. We test the following implementations of the iterated Prisoner's dilemma with two agents.

\begin{figure*}
     \subfloat[Rewards \label{fig:ipd_basic_implementations}]{%
       \includegraphics[width=0.32\textwidth]{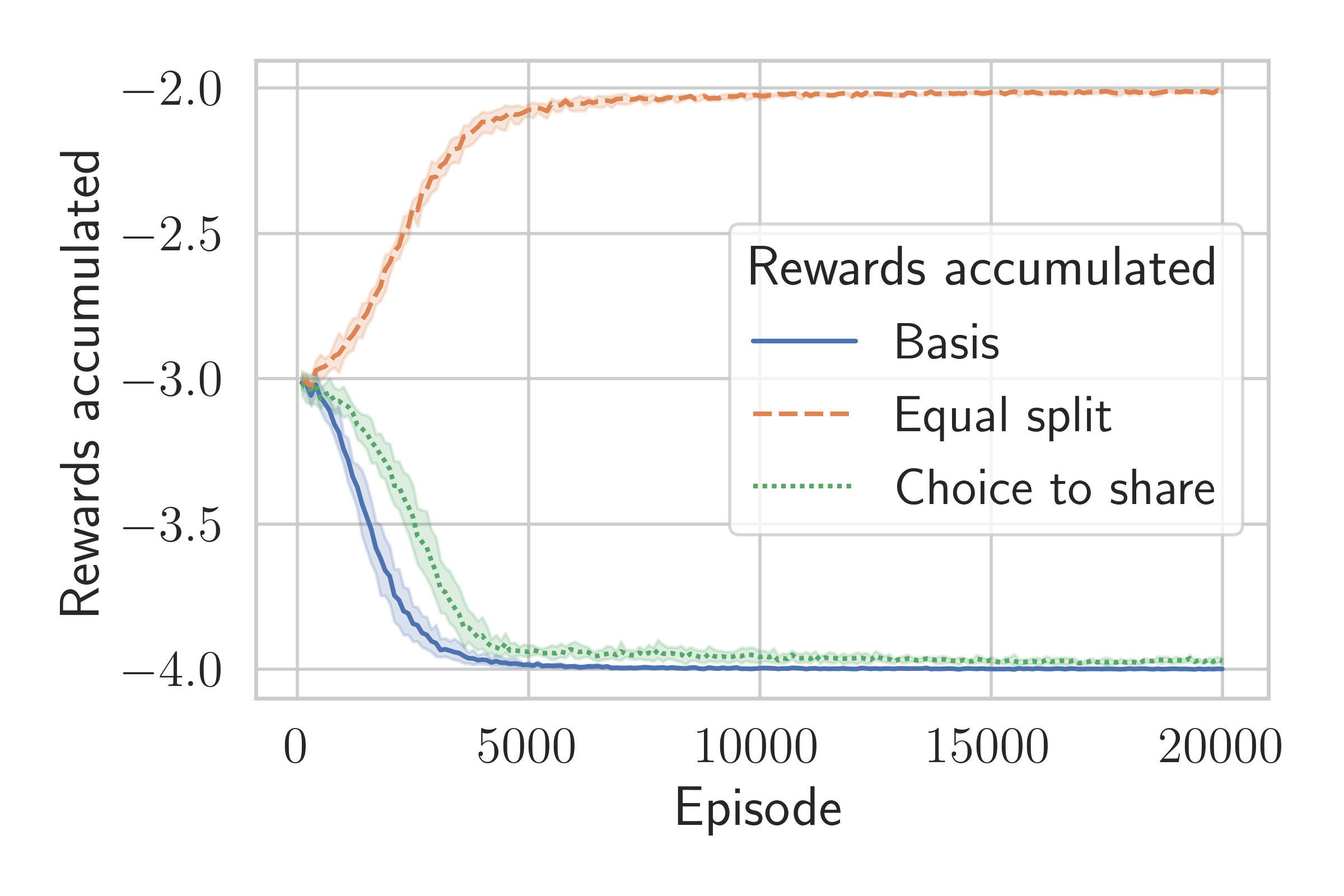}
     }
     \hfill
     \subfloat[Rewards with trading of 10\% or 50\% shares \label{fig:ipd_trading-1}]{%
       \includegraphics[width=0.32\textwidth]{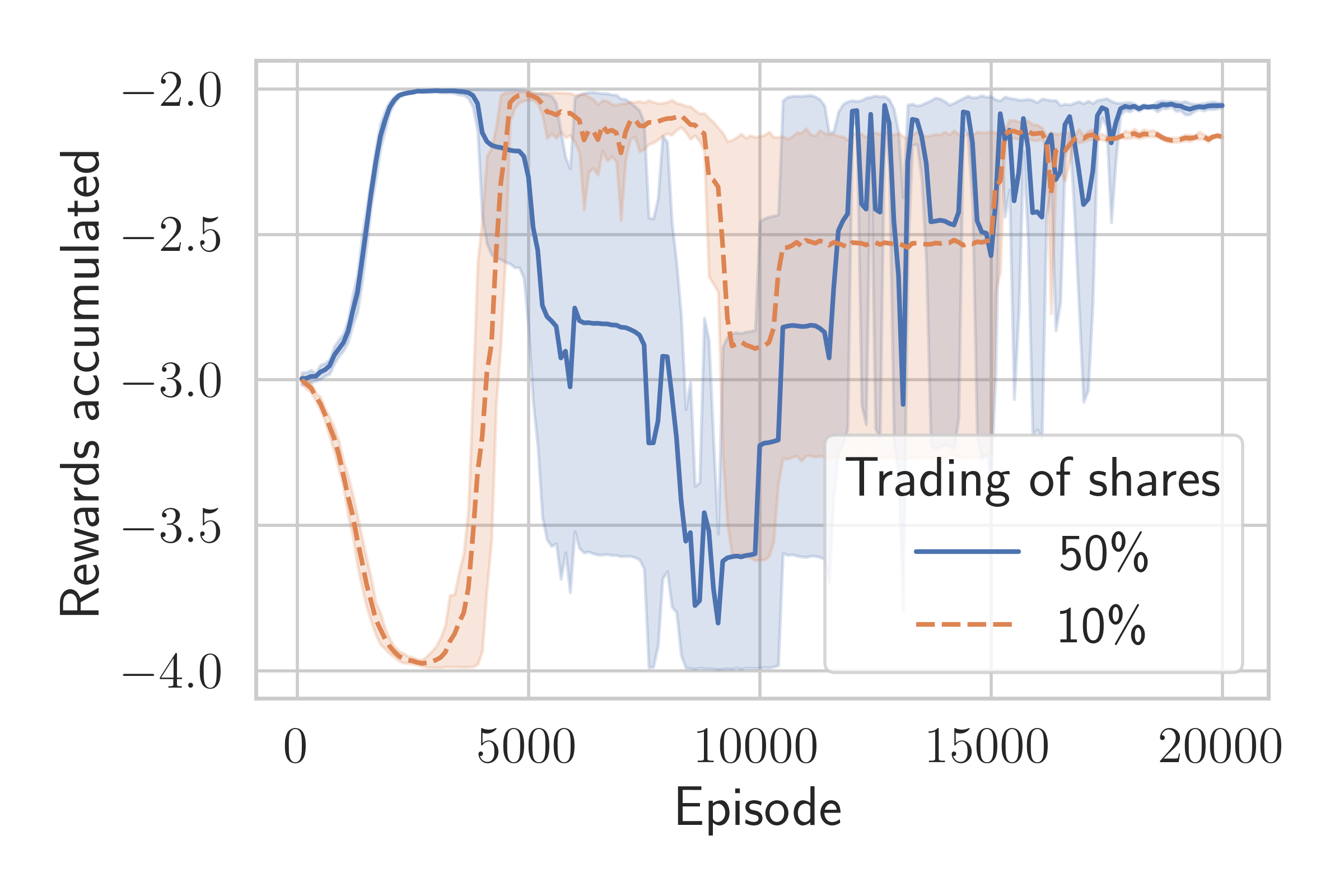}
     }
     \hfill
     \subfloat[Shares in own reward \label{fig:ipd_trading-2}]{%
       \includegraphics[width=0.32\textwidth]{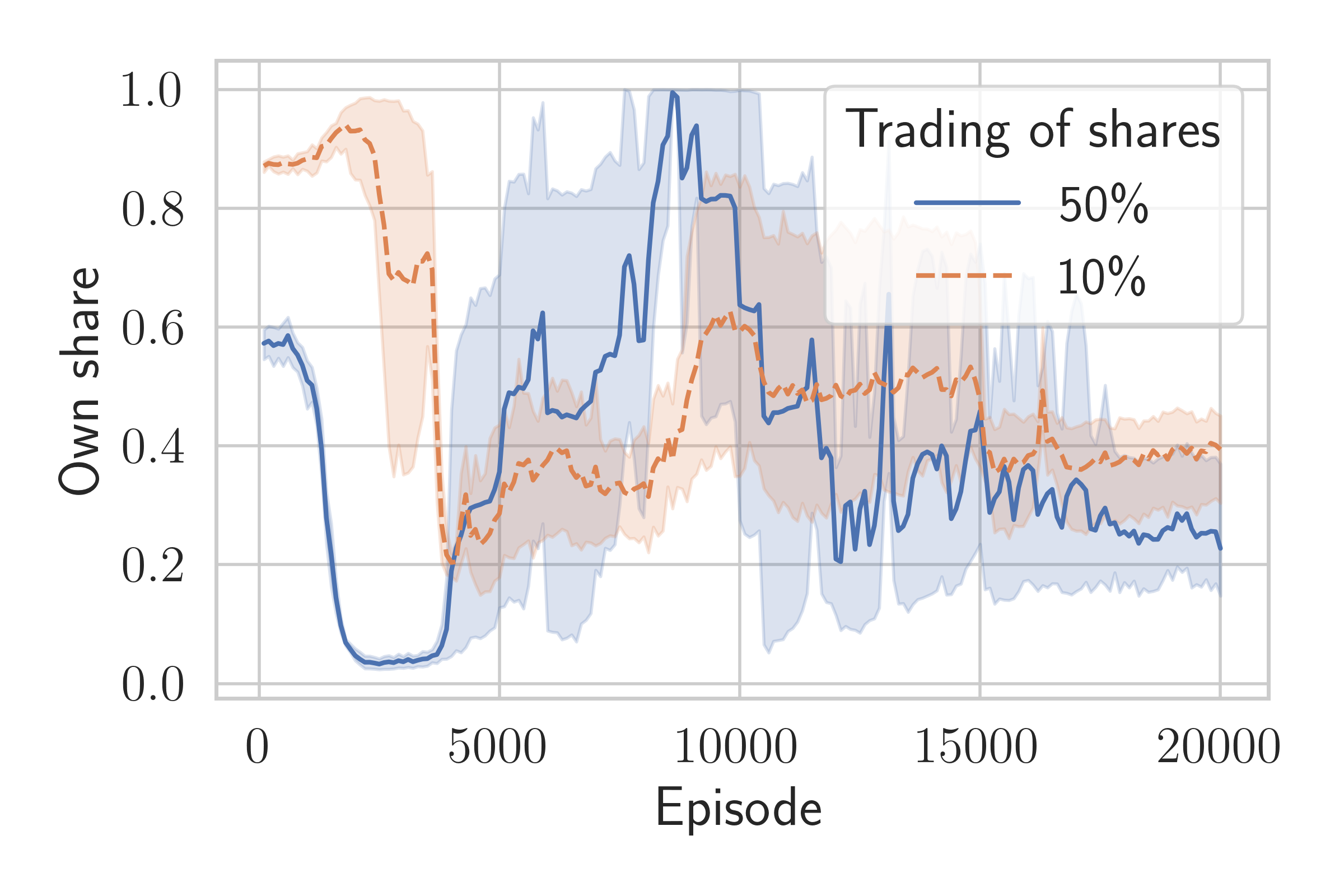}
     }
     \caption{Results from the iterated Prisoner's dilemma with and without participation.}
     \label{fig:ipd-results}
\end{figure*}

\textbf{(\romannumeral 1) No participation:} The possible environment actions of each agent are \textit{cooperation} and \textit{defection}. In the implementation, these two actions are encoded as 0 and 1. There are four possible states:  (cooperation, cooperation), (cooperation, defection), (defection, cooperation), (defection, defection). If the agents chose their actions randomly, the average individual reward would be $(-1+0-3-2)/4=-1.5$. If instead the agents learned to maintain a cooperative equilibrium in which both choose \textit{cooperation}, the individual rewards would converge to $-1$. But according to theory, \textit{defection} is a dominant action in the ``classical'' Prisoner's dilemma. In that case, the individual rewards would converge to $-2$.  Indeed, in the implementation without participation shares, a stable equilibrium evolves in which both agents \textit{defect}. The accumulated reward converges to $-2+(-2)=-4$ (\autoref{fig:ipd_basic_implementations}).

\textbf{(\romannumeral 2) Equal distribution of individual rewards:} In this implementation, the agents always receive the average reward of all individual shares. Mathematically speaking, this means that all individual rewards are aggregated and then divided by the number of agents: $\frac{1}{n} \times \sum_i^n reward_i$ for $n$ agents. The action space and everything else stays the same. An agent cannot choose between sharing the rewards or receiving the individual reward. Each agent learns to \textit{cooperate}, which leads to an accumulated reward of $-1+(-1)=-2$. In this implementation, cooperating is a dominant action. This implementation demonstrates that the sharing of rewards can lead to cooperation. However, it does not demonstrate whether the agents can actively find and maintain such an equilibrium, when they can freely trade (\autoref{fig:ipd_basic_implementations}).

\textbf{(\romannumeral 3) Choosing whether to share rewards:} The agents can decide to share their individual rewards. The action space is extended. In addition to the two environment actions, an agent can choose between sharing the rewards or receiving the individual reward. The action space is defined by an action tuple: the environment action and the trade action. Hence, there are now $2 \times 2 = 4$ possible actions per agent. Importantly, an individual agent cannot determine whether the combined rewards are evenly divided between both agents. Instead, this is only the case if both agents decide to share their rewards. The state space is extended to eight states: $\textit{sharing} \times \textit{env. action 1} \times  \textit{env. action 2} = 2 \times 2 \times 2 = 8$. Both agents learn to \textit{defect}, which leads to an accumulated reward of $-2+(-2)=-4$ (\autoref{fig:ipd_basic_implementations}). In this implementation, defecting without sharing is a dominant action. Hence, the sole opportunity to share rewards is not enough.

\textbf{(\romannumeral 4) Trading 50\% shares:} Initially, both agents do not hold participation shares of the other agent. In every step, they can choose between six actions. An action can be regarded as a 3-tuple: the environment action, whether to increase the shares of the own rewards, and whether to increase the shares of the other agent's rewards. When an agent decides to cooperate, there are three possible actions: (cooperate, not buy own shares, not buy other agent's shares), (cooperate, buy own shares, not buy other agent's shares), (cooperate, not buy own shares, buy other agent's shares). Symmetrically, there are three possible actions for defection. Again, a trade of shares is only executed if both agents intend to do so. For instance, if both choose to buy own shares and they hold 50\% in each agent's rewards, they exchange shares and now hold 100\% of their own shares. This would mean that there is no participation anymore. However, if they now choose to buy own shares again, this trade cannot occur, as they already hold all of their own shares. As a consequence, only the environment action has an effect. The implication is that all shares are valued at the same price, they are just exchanged in the same proportions, and the sum of shares per agent is always 100\%. There are 36 states: $\textit{env. actions} \times \textit{portion own shares} \times \textit{trade} = 4 \times 3 \times 3$. The portion of own shares can be 0, 0.5, or 1. The ``trade'' variable represents whether there is no trade, a trade which leads to an increasing amount of own shares, or a trade which leads to a decreasing amount of own shares. This implementation is successful in establishing cooperation. The accumulated rewards converge to $-1+(-1)=-2$  (\autoref{fig:ipd_trading-1}). However, the learning process takes rather long. Additionally, the actions and rewards first drift off to the previous inefficient equilibrium.

\textbf{(\romannumeral 5) Trading 10\% shares:} Compared to the trading of 50\% shares, the state space increases to $4 \times 11 \times 3 = 132$ because the proportion of own shares can now be 0, 0.1, 0.2, ..., 1. Again, a socially optimal equilibrium can be established. A proportion of around 40\% own shares and 60\% other shares seems to be efficient to make the agents not deviate to defection (\autoref{fig:ipd_trading-2}).

\subsection{The clean-up game}

In the clean-up game, multiple agents simultaneously attempt to collect apples in the same environment. An agent gets rewarded +1 for each apple that they collect. The apples spawn randomly on the right side of a quadratic \textit{7x7} or \textit{10x10} map. On the left side of the map, there is a river that gets increasingly polluted with waste. 
As the waste level increases and approaches a depletion threshold, the apple spawn rate decreases linearly to zero. To avoid a quick end of the game, an agent can fire a \textit{cleaning beam} to clear waste. But an agent can only do so when being in the river. The cleaning beam then clears all the waste upwards from the agent. Each agent's observation is an egocentric RGB image of the whole map. The dilemma is that clearing waste by staying in the river and firing cleaning beams is less attractive than receiving rewards by collecting apples. However, if all agents focus on only collecting apples, the game is quickly over and the total reward of the agents remains very low. Hence, the game is an \textit{intertemporal} social dilemma, in which there is a trade-off between short-term individual incentives and long-term collective interest \cite{hughes2018inequity}. This domain is especially interesting as models based on behavioral economics can only explain cooperation in simple, unrealistic, stateless matrix games. In contrast, the cleanup game is a temporally and spatially extended Markov game. 

\subsection{Participation in the clean-up game with two agents}

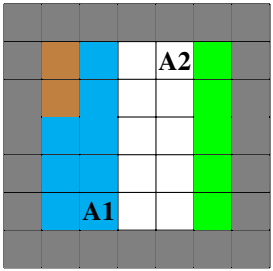
\begin{figure}[!htb]
\centering
		\vspace{0pt}
		\hspace{0pt}
            \centering
\begin{tikzpicture}[every node/.style={minimum size=.5cm -\pgflinewidth, outer sep=0pt}] 
    \draw[step=0.5cm,color=black] (-1,-1) grid (2.5,2.5);
    \foreach \x/\y in {-0.75/+2.25, -0.25/+2.25, +0.25/+2.25, +0.75/+2.25, +1.25/+2.25, +1.75/+2.25, +2.25/+2.25,
    				-0.75/-0.75, -0.25/-0.75, +0.25/-0.75, +0.75/-0.75, +1.25/-0.75, +1.75/-0.75, +2.25/-0.75,
    				-0.75/-0.25, -0.75/+0.25, -0.75/+0.75, -0.75/+1.25, -0.75/+1.75,
    				+2.25/-0.25, +2.25/+0.25, +2.25/+0.75, +2.25/+1.25, +2.25/+1.75}
    		\node[fill=gray] at (\x,\y) {};
    \node at (1.25,+1.75) {\textbf{A2}};
    \node[fill=green] at (1.75,+1.75) {};
    \node[fill=green] at (1.75,+1.25) {};
    \node[fill=green] at (1.75,+0.75) {};
    \node[fill=green] at (1.75,+0.25) {};
    \node[fill=green] at (1.75,-0.25) {};
    \node[fill=brown] at (-0.25,+1.75) {}; \node[fill=cyan] at (+0.25,+1.75) {};
    \node[fill=brown] at (-0.25,+1.25) {}; \node[fill=cyan] at (+0.25,+1.25) {};
    \node[fill=cyan] at (-0.25,+0.75) {}; \node[fill=cyan] at (+0.25,+0.75) {};
    \node[fill=cyan] at (-0.25,+0.25) {}; \node[fill=cyan] at (+0.25,+0.25) {};
    \node[fill=cyan] at (-0.25,-0.25) {}; \node[fill=cyan] at (+0.25,-0.25) {};
    \node at (+0.25,-0.25) {\textbf{A1}};
\end{tikzpicture}
\caption[Small Cleanup map]{Small \textit{7x7} map with two agents. Agent 1 gets spawned by the river, whereas agent 2 gets spawned close to the apples. If they do not divide their tasks, the game is likely to stop soon, as the waste expands downwards if it is not cleared by the agents.}
\label{fig:cleanup_map}
\end{figure}

The smaller \textit{7x7} map is used as described in \cite{lio}. The initial state of the game is displayed in \autoref{fig:cleanup_map}. 
The blue cells on the left side of the map represent the river, the green cells indicate the area where apples are randomly spawned, and the brown cells represent the waste.
Agent 1 gets spawned in the river, which enables it to clear the waste. 
Agent 2 gets spawned close to the green area on the right where apples are randomly spawned. 
This already gives a hint at one possible strategy, which consists of agent 1 clearing the waste, and agent 2 collecting the apples. However, there is no justifying reason for agent 1 to clear waste as long as it does not get the chance to collect apples or profits from agent 2's rewards. Agent 2 is in a similar dilemma. If it attempts to collect apples by being in the green area, it cannot fire the cleaning beam. Without a division of work between the agents, the game must stop early, as the blue and green area are too far away. The possible environment actions of each agent are \textit{move left}, \textit{move right}, \textit{move up}, \textit{move down},  \textit{no operation}, and \textit{cleaning}. 
In all implementations of the clean-up domain, an actor-critic method is used. The optimization is decentralized. 

We use the same convolutional networks to process image observations in Cleanup as \textit{LIO}. The policy network has a softmax output for discrete actions in all environments. For all experiments, We use the same neural architecture. We test the following implementations of the clean-up game with two agents.

\begin{figure*}
\subfloat[Two agents]{%
\includegraphics[width=0.48\textwidth]{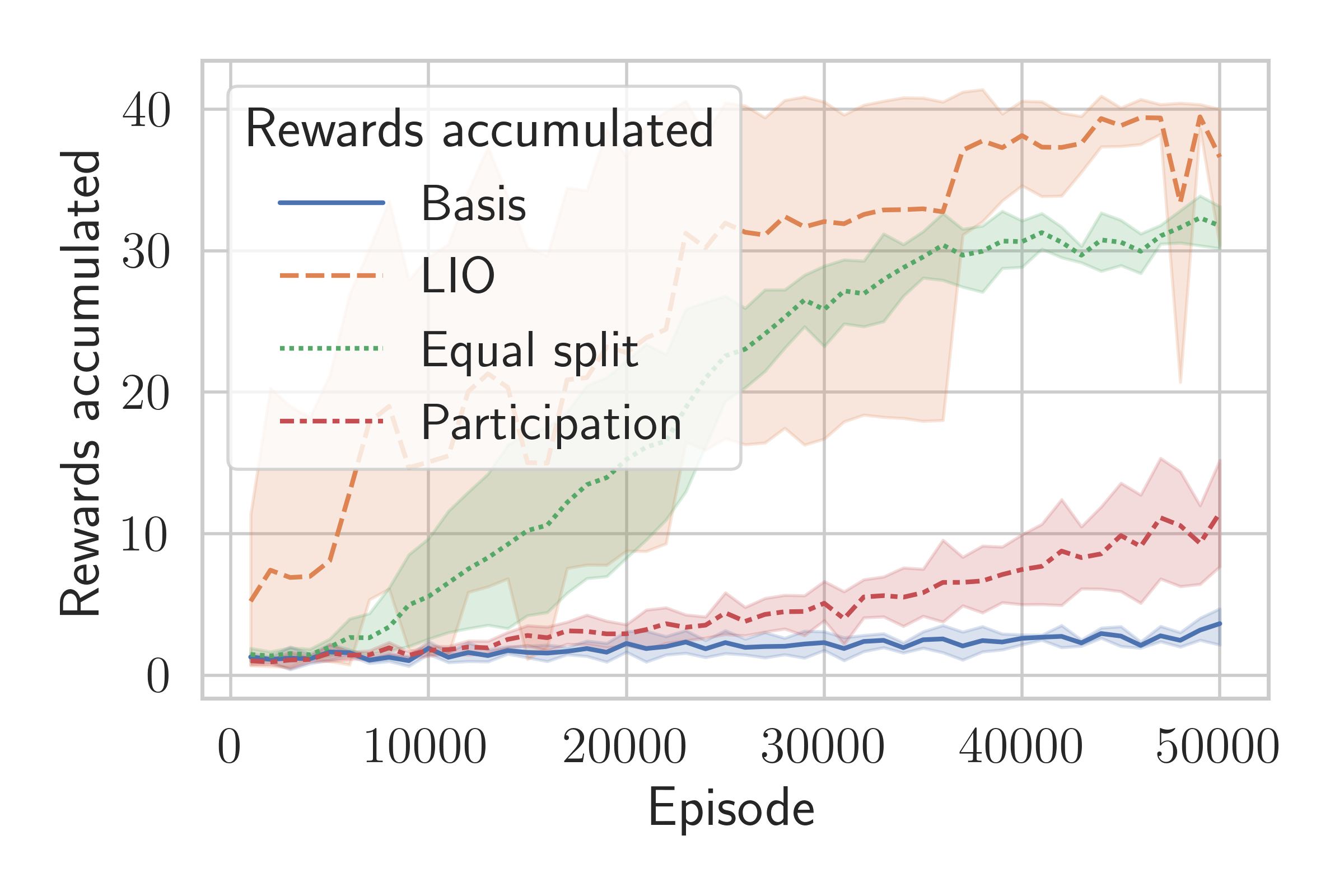}
\label{fig:cleanup_n2_rewards}
}
\subfloat[Three agents]{%
\includegraphics[width=0.48\textwidth]{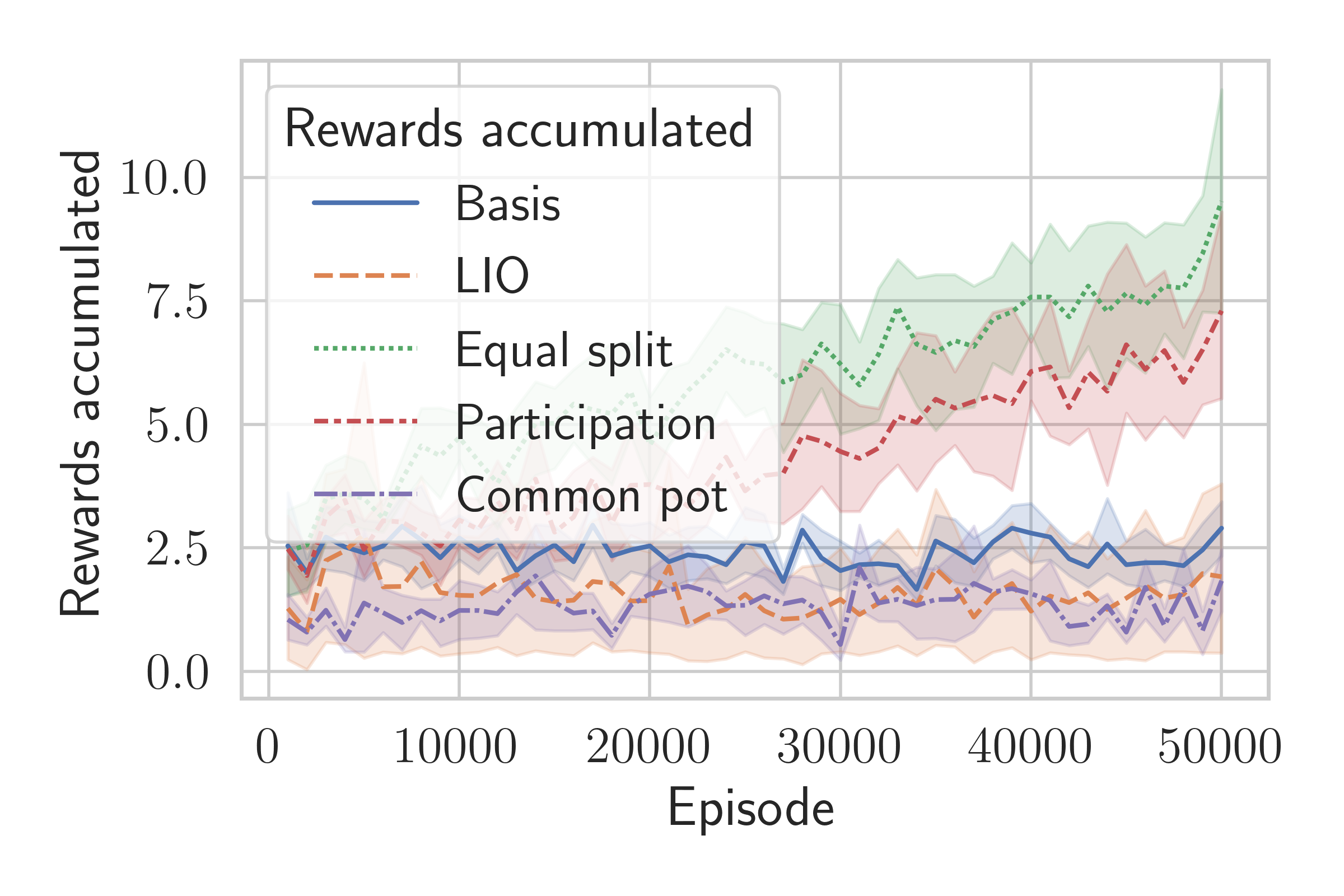}
\label{fig:cleanup_n3_rewards}
}
\hfill
\caption{Results from the clean-up game}
\label{fig:clean-up-results}
\end{figure*}

\textbf{(\romannumeral 1) No participation:} Without any additional incentive structures, the social dilemma seems unsolvable. As depicted in \autoref{fig:cleanup_n2_rewards}, the accumulated reward remains at a very low level. Cooperation is not attractive for any of the agents. 
It is difficult to learn that clearing waste is creating value since the apples are spawned far away from the river. If the agent moves back from the river to the apples, the game is likely to stop before it even collects an apple. 

\textbf{(\romannumeral 2) LIO:} The baseline scenario is augmented with the possibility for each agent to give rewards to the other agent as an additional channel for cooperation. Importantly, the additional reward payments $r^j_{\eta^i}$ from agent $i$ to agent $j$ is learned via direct gradient ascent on the agent’s own extrinsic objective, involving its effect on all other agents’ policies. Hence, the payments are not part of the reinforcement learning, leaving the action space of the reinforcement learner unaffected (instead of augmenting it). The agents successfully divide their tasks. The one that is closer to the river and waste becomes the cleaner, whereas the one that is closer to the apples specializes in collecting apples. See \cite{lio} for implementation details.

\textbf{(\romannumeral 3) Equal distribution of individual rewards:} In this scenario, the baseline scenario is augmented with the equal distribution of the joint rewards between the agents after each step. This additional computation does not affect the state or action space. Similar to \textit{LIO}, the one that is closer to the river turns into the cleaner, whereas the one that is closer to the apples specializes in collecting apples. Both profit from this task division, as both are evenly rewarded for any apple being collected by anyone. In comparison with \textit{LIO}, the learning process looks less volatile. 

\textbf{(\romannumeral 4) Pre-trade of participation rights:} In this scenario, there is an additional first step added to each episode of the baseline scenario. In this first step, both agents can choose between six actions (0-5), representing $0\%$, $20\%$, $40\%$, $60\%$, $80\%$, and $100\%$. The maximum of both agents' chosen number determines the participation in their own rewards over the episode. The remaining portion is the participation in the other agent's individual rewards. For instance, if agent 1 chooses 40\%, and agent 2 prefers 80\%, they will receive $80\%$ of their individual rewards and $20\%$ of the other agent's individual rewards. 
The additional first step is part of the reinforcement learning, but no rewards are distributed in this step.
The idea behind the additional trade step is that both agents can avoid cooperation by choosing $100\%$. By taking the maximum, the more conservative action is executed. 
Again, the agents successfully divide their tasks. The amount of waste cleared by agent 1 moves in parallel with the accumulated rewards. However, the magnitude of the joint rewards only reaches around 11, and it is unclear whether this is a stable level. 


\begin{figure*}[!htb]
        \centering
        \begin{subfigure}[b]{0.3\textwidth}
            \centering
            \includegraphics[width=\textwidth]{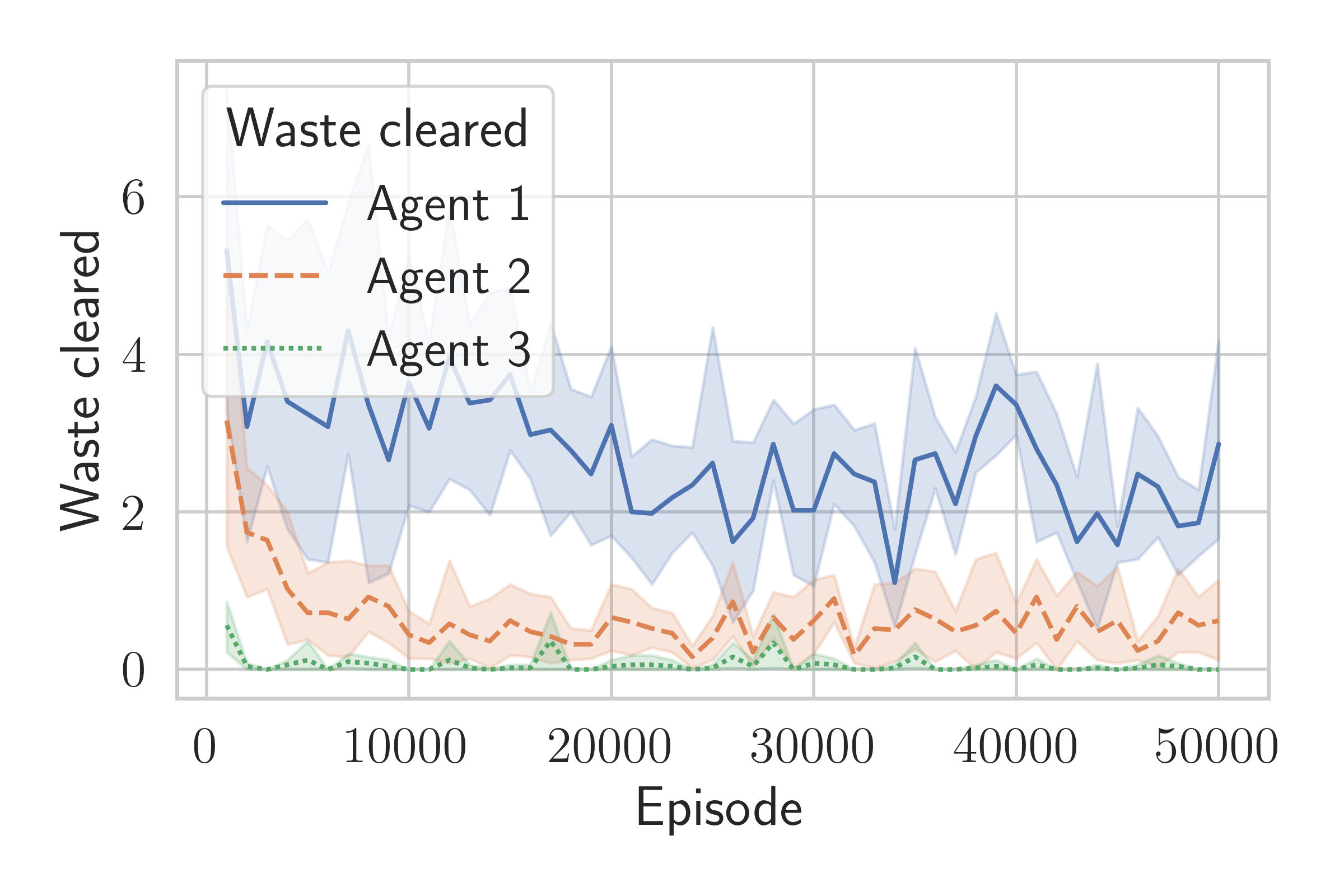}
        \end{subfigure}
        \hfill
        \begin{subfigure}[b]{0.3\textwidth}  
            \centering 
            \includegraphics[width=\textwidth]{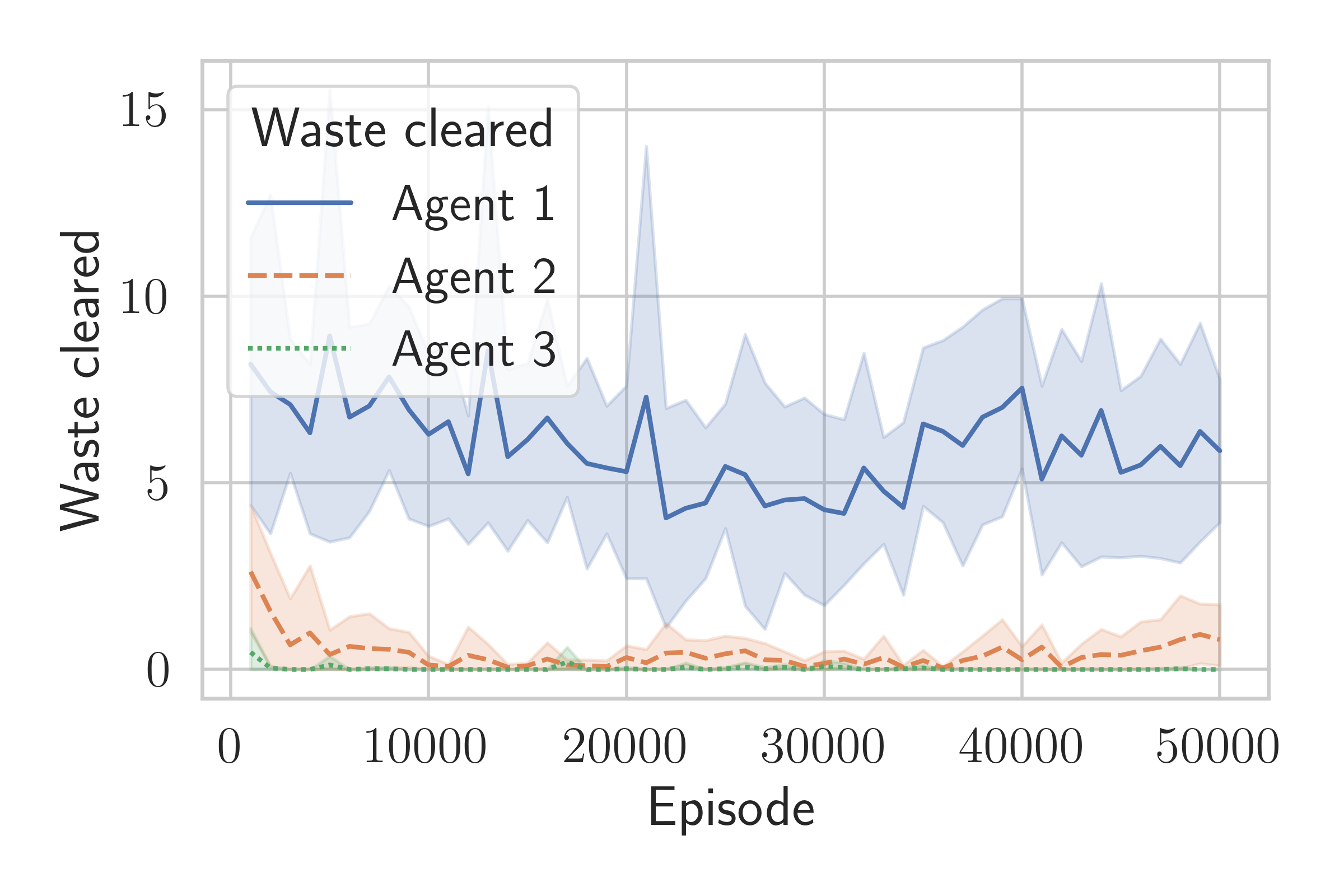}
             \end{subfigure}
        \hfill
        \begin{subfigure}[b]{0.3\textwidth}  
            \centering 
            \includegraphics[width=\textwidth]{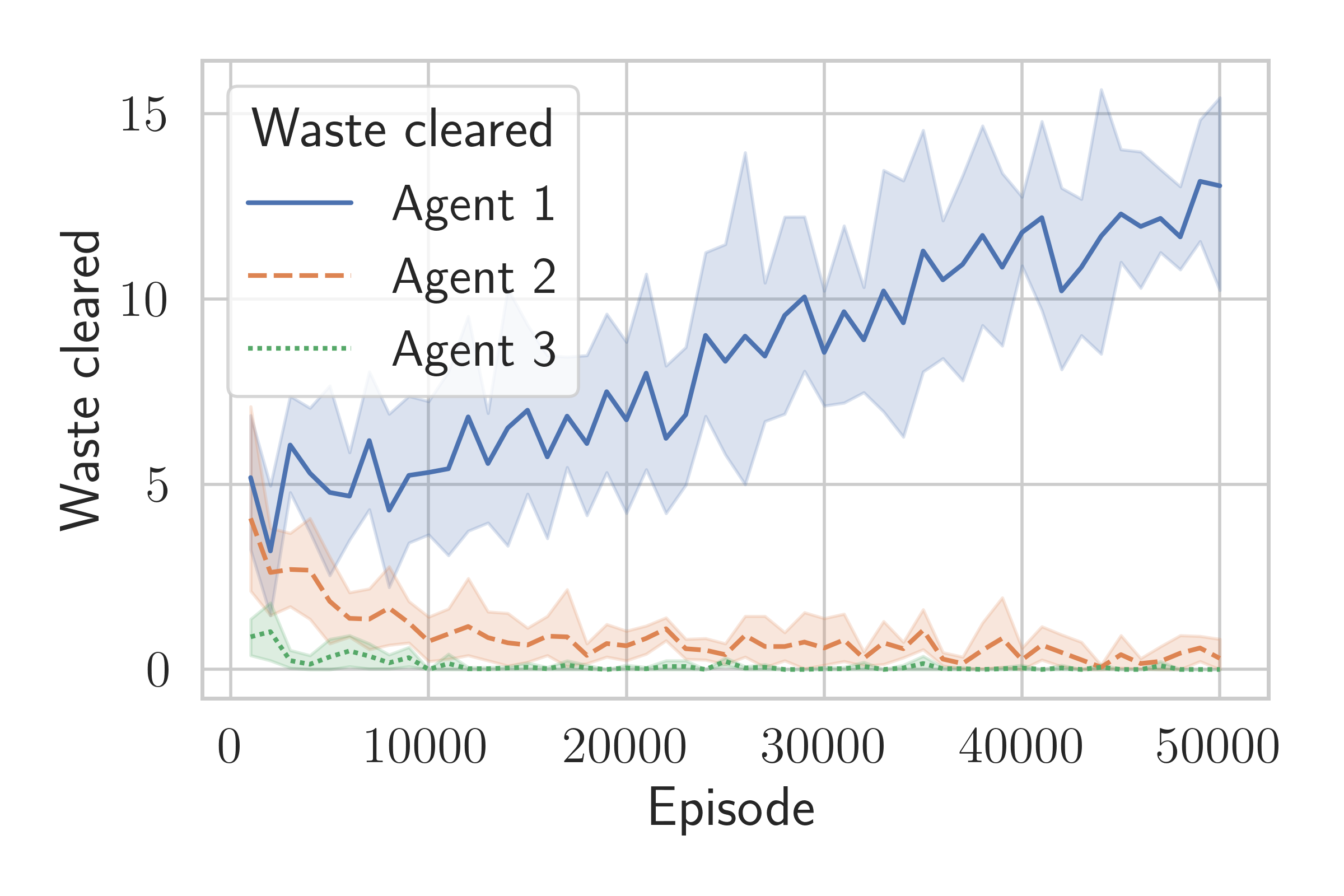}
        \end{subfigure}
        \vskip\baselineskip
        \begin{subfigure}[b]{0.3\textwidth}   
            \centering 
            \includegraphics[width=\textwidth]{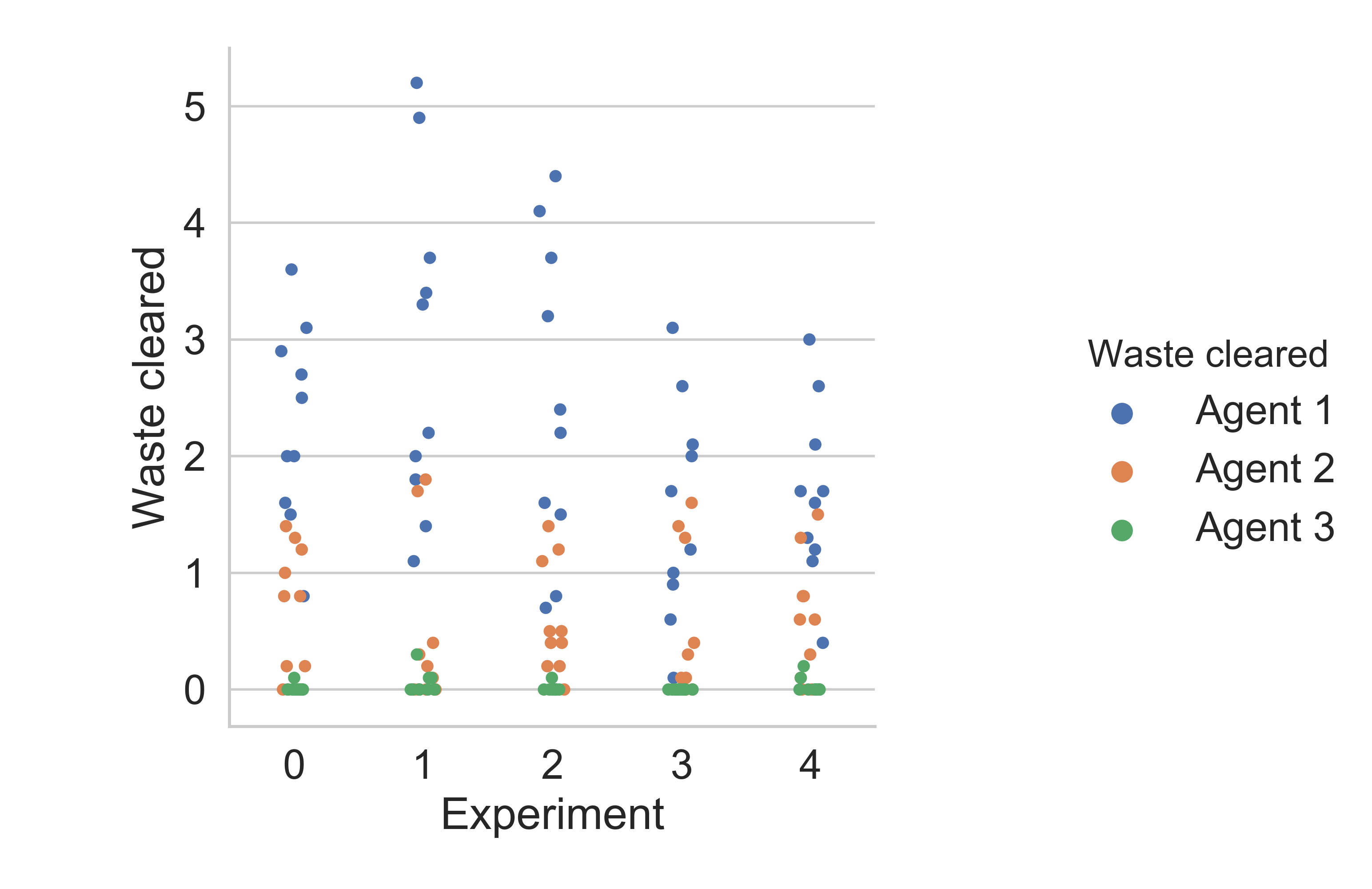}
            \caption[]%
            {{\small In the basic implementation, none of the agents learns to clear waste.}}   
            \label{fig:cleanup_n3_waste_basis}
        \end{subfigure}
        \hfill
        \begin{subfigure}[b]{0.3\textwidth}   
            \centering 
            \includegraphics[width=\textwidth]{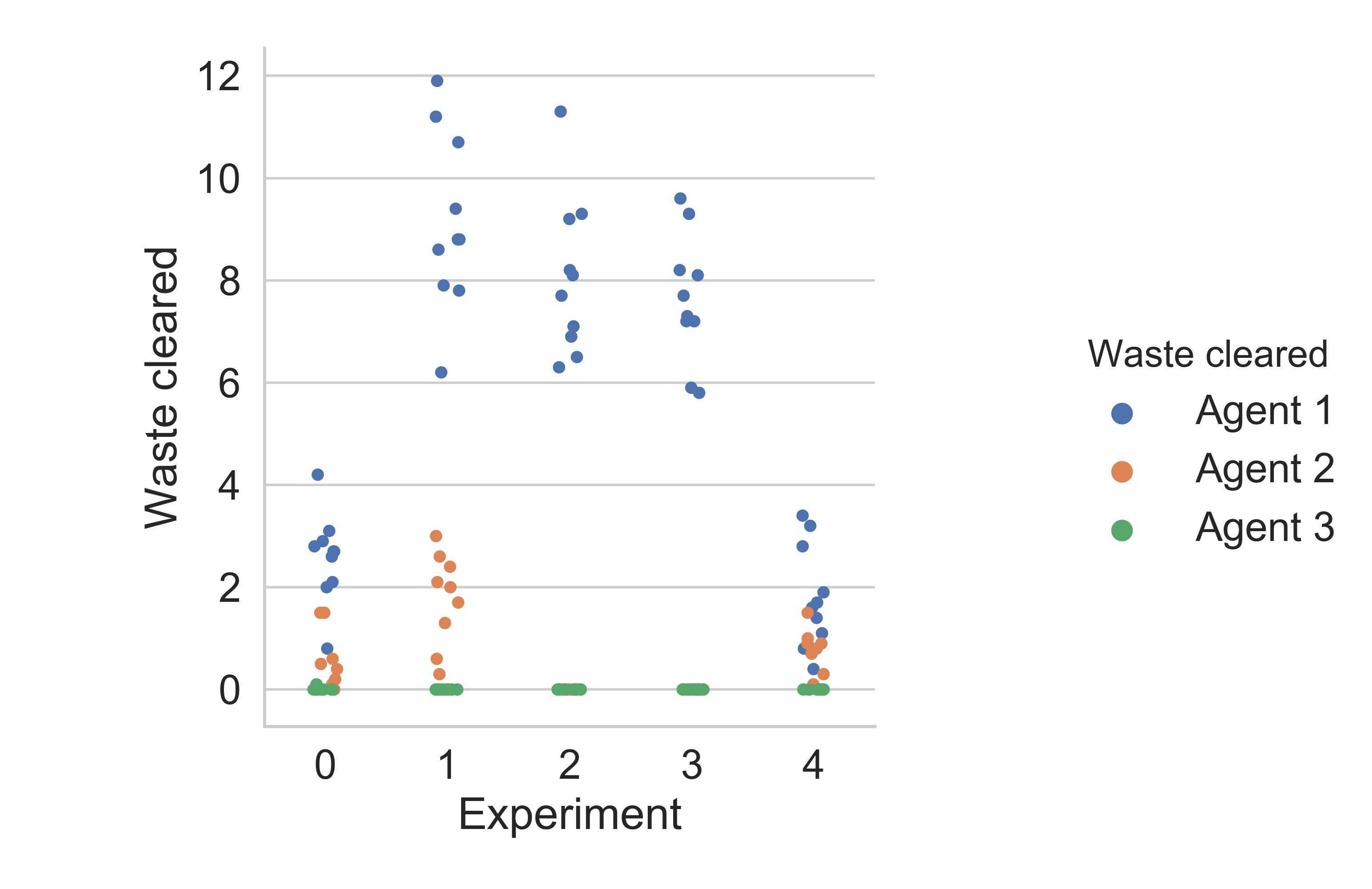}
            \caption[]%
            {{\small In the LIO implementation, no clear strategy evolves. 
            }}
            \label{fig:cleanup_n3_waste_lio}
            \end{subfigure}
        \hfill
        \begin{subfigure}[b]{0.3\textwidth}   
            \centering 
            \includegraphics[width=\textwidth]{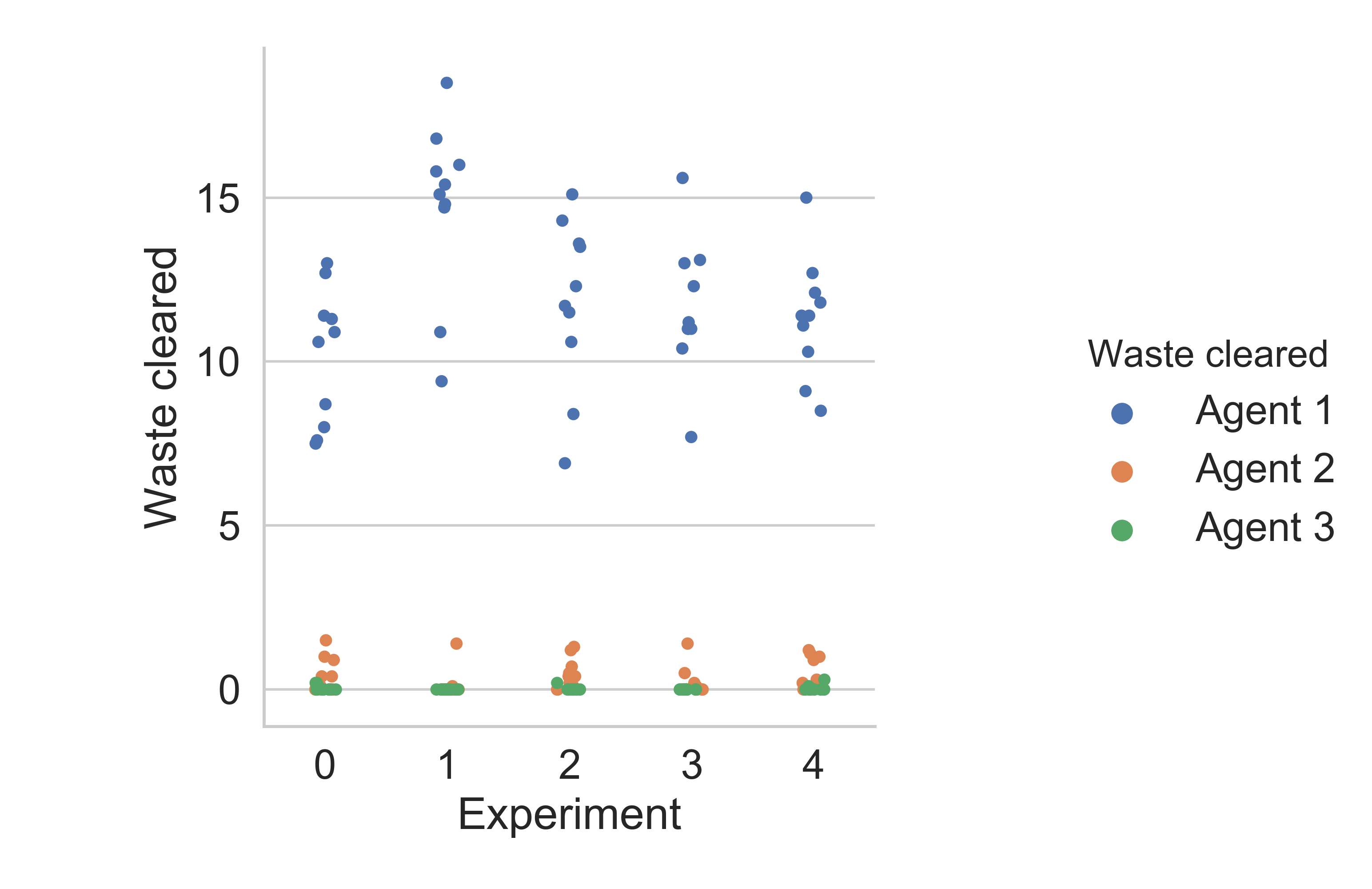}
            \caption[]%
            {{\small With participation, one of the agents learns to clear waste.}}    
            \label{fig:cleanup_n3_waste_participate}
        \end{subfigure}
        \caption[]
        {\small Waste cleared in clean-up with three agents for the basic implementation, LIO, and participation.} 
        \label{fig:cleanup_n3_waste}
\end{figure*}

\section{Conclusion}
\label{sec:conclusion}

By introducing the idea of participating in other agents' rewards, we suggest a new method for coordination and cooperation in shared multi-agent environments. 
Agents learn that direct participation in other agents' rewards reinforces policies that optimize a global objective. Through this mechanism, no additional extrinsic incentive structures are needed such as in \textit{LIO} \cite{lio}. Other previous works focused on intrinsic incentives \cite{eccles2019learning,hughes2018inequity,wang2018evolving} which would not be needed either. Especially the simplistic and graspable extension of standard models makes the participation appealing. 
In the two tested social dilemma problems, the Iterated Prisoner’s dilemma and Cleanup, the opportunity to participate via shares is used by the agents to discover cooperative behaviors. In fact, the division of labor in Cleanup is effectively enabled. Without the participation, the agents cannot learn to divide their task into subtasks, as clearing waste does not directly lead to rewards. But through the participation, the positive impact of clearing waste is directly observed through the other agents' rewards from collecting apples. Importantly, the other agent can only collect apples thanks to the clearing of the waste beforehand. The introduced method of participation can achieve optimal collective performance in the Prisoner's dilemma. 
In Cleanup, the improvement in the collective performance through participation is clearly visible. Although it is not clear whether the agents exhaustively learn the perfect participation allocation, the agents manage to coordinate and enhance their performance over time.

Our approach attempts to answer some open questions regarding cooperation in a decentralized multi-agent population. Firstly, although the agents must simultaneously learn how to participate via shares in addition to learning their environment actions, learning to keep shares of other agents may be a simpler task than punishing other agents whenever they act only in their own interest. 
Secondly, punishments or other incentivizing rewards should be earned beforehand. When working with shares, they can be directly traded without costs. Thirdly, a share market can be implemented in various ways. We are certain that a suitable market structure can be found in most social dilemmas. 
Furthermore, \textit{LIO} assumes that recipients cannot reject an incentive, but an agent may accept only some incentives. In the case of shares, this is not a problem as the respective rewards are just distributed according to the owned shares. Another benefit of the participation approach is that there is no clear strategy for the agents to misuse the additional market feature to exploit the other agents. There is an emerging literature on reward tampering \cite{everitt2021reward}, and a participation market could be a step in the right direction of deploying safe applications.

Our work contributes to the aim of ensuring the common good in environments with independent agents. Although the participation approach works well in the tested social dilemmas, it remains unclear whether this is also the case in other environments. Another open question is if the agents can always find a stable allocation of shares. We suggest experiments with a broker that sets prices in combination with a limit order book for matching demand and supply. Additionally, participation needs to be tested in other environments with more agents as well as other game structures.

\vfill

\bibliographystyle{apalike}
{\small
\bibliography{main}}

\end{document}